\documentclass[twocolumn,showpacs,preprintnumbers,amsmath,amssymb,floatfix,superscriptaddress]{revtex4}

\usepackage[dvips]{graphicx}
\usepackage{dcolumn}
\usepackage{bm}
\usepackage{amsmath,amsthm,amssymb}
\usepackage{graphicx}
\usepackage{epstopdf}
\usepackage{psfrag}
\usepackage{color}
\usepackage[ansinew]{inputenc}
\usepackage{rotating}
\usepackage{ulem}
\def\beq{\begin{equation}}
\def\eeq{\end{equation}}

\newcommand{\ket}[1]{|{#1}\rangle}

\newcommand{\affA}{%
    Institut f{\"u}r Physik, Johannes Gutenberg-Universit\"at Mainz, D-55099 Mainz, Germany}
\newcommand{\affB}{%
Zentrum f\"ur Optische Quantentechnologien and The Hamburg Centre for Ultrafast Imaging,  Universit\"at Hamburg, Luruper Chaussee 149, D-22761 Hamburg, Germany}

\begin{document}

\preprint{}

\title{Quantum dynamics of an atomic double-well system interacting with a trapped ion}

\author{J. Joger}\affiliation{\affA}
\author{A. Negretti}\affiliation{\affB}
\author{R. Gerritsma}\affiliation{\affA}

\email{rene.gerritsma@uni-mainz.de}
\affiliation{}\homepage{http://www.science.uva.nl/research/aplp/}

\date{\today}

\begin{abstract}
We theoretically analyze the dynamics of an atomic double-well system with a single ion trapped in its center. We find that the atomic tunnelling rate between the wells depends both on the spin of the ion via the short-range spin-dependent atom-ion scattering length and on its motional state with tunnelling rates reaching hundreds of Hz. A protocol is presented that could transport an atom from one well to the other depending on the motional (Fock) state of the ion within a few ms. This phonon-atom coupling is of interest for creating atom-ion entangled states and may form a building block in constructing a hybrid atom-ion quantum simulator. We also analyze the effect of imperfect ground state cooling of the ion and the role of micromotion when the ion is trapped in a Paul trap. Due to the strong non-linearities in the atom-ion interaction, the micromotion can cause couplings to high energy atom-ion scattering states, preventing accurate state preparation and complicating the double-well dynamics. We conclude that the effects of micromotion can be reduced by choosing ion/atom combinations with a large mass ratio and by choosing large inter-well distances. The proposed double-well system may be realised in an experiment by combining either optical traps or magnetic microtraps for atoms with ion trapping technology. 
\end{abstract}



\maketitle

Recent experiments involving a combination of ultra-cold quantum gases and trapped ions have sparked significant interest in studying their properties in the quantum regime~\cite{Grier:2009,Zipkes:2010,Schmid:2010,Zipkes:2010b,Harter:2013}. These experiments explore sympathetic cooling of ions by means of clouds of cold atoms~\cite{Krych:2010} and study cold chemistry~\cite{Rellergert:2011}. Trapped single ions may be used to perform in situ measurements of cold atomic gases in lattice potentials~\cite{Kollath:2007}, or to study the physics of impurities in one-dimensional Bose gases~\cite{Goold:2010}. The excellent controllability of trapped ions together with the near perfect state-preparation and read-out may also allow experiments in which ions {\it  control} the dynamics of ultra-cold atoms. For instance, it has been proposed that an entangling quantum gate operation could be performed on a single trapped atom and ion by means of a controlled collision~\cite{Doerk:2010}. Here, the spin dependence of the scattering cross section can be used to obtain a state-dependent collisional phase shift, leading to the desired quantum logic~\cite{Calarco:2004,Krych:2009}. In a recent paper we have shown that a similar controllability should arise when a single trapped ion is placed in between a double-well atomic Josephson junction. Here, the spin of the ion would control the tunnelling rate of atoms between the wells via the state-dependent short-range interactions of the atoms and ion~\cite{Gerritsma:2012}. Since the ion could in this way control many-body dynamics, mesoscopic entanglement between the atomic matter wave and the spin of the ion may be created. The interplay between the spin-dependent tunnelling and the inter-atomic interactions could also result in superpositions of quantum self-trapping~\cite{Albiez:2005} and Josephson tunnelling. Since trapped ions allow for superb experimental control, they may be better suited to investigate Josephson physics than using single atomic impurities~\cite{Bausmerth:2007,Fischer:2007}. 

 Including a single trapped ion expands on the rich dynamics of the Josephson junction as described in numerous experimental and theoretical works~\cite{Albiez:2005,Gati:2007,Levy:2007,LeBlanc:2011,Bouchoule:2005,Betz:2011,Eckardt:2005,Grond:2011,Bar-Gill:2011,He:2011}. Furthermore, the system may be seen as a `unit cell' for a larger scale hybrid atom-ion quantum simulator~\cite{Bissbort:2013}. Constructing such a device by concatenating ion-controlled double-wells could be a natural way to combine quantum simulators in which atoms can tunnel between sites in an optical lattice~\cite{Bloch:2012} with  simulators employing the pseudo-spin and collective motional states of ion crystals~\cite{Blatt:2012}. In such a system, atomic Bloch waves would interact with phononic excitations in the ion crystal, leading to solid state phenomena such as Peierls instabilities~\cite{Bissbort:2013, Lan:2013} and phonon-mediated interactions. 

\begin{figure}[b]
\includegraphics[width=6cm]{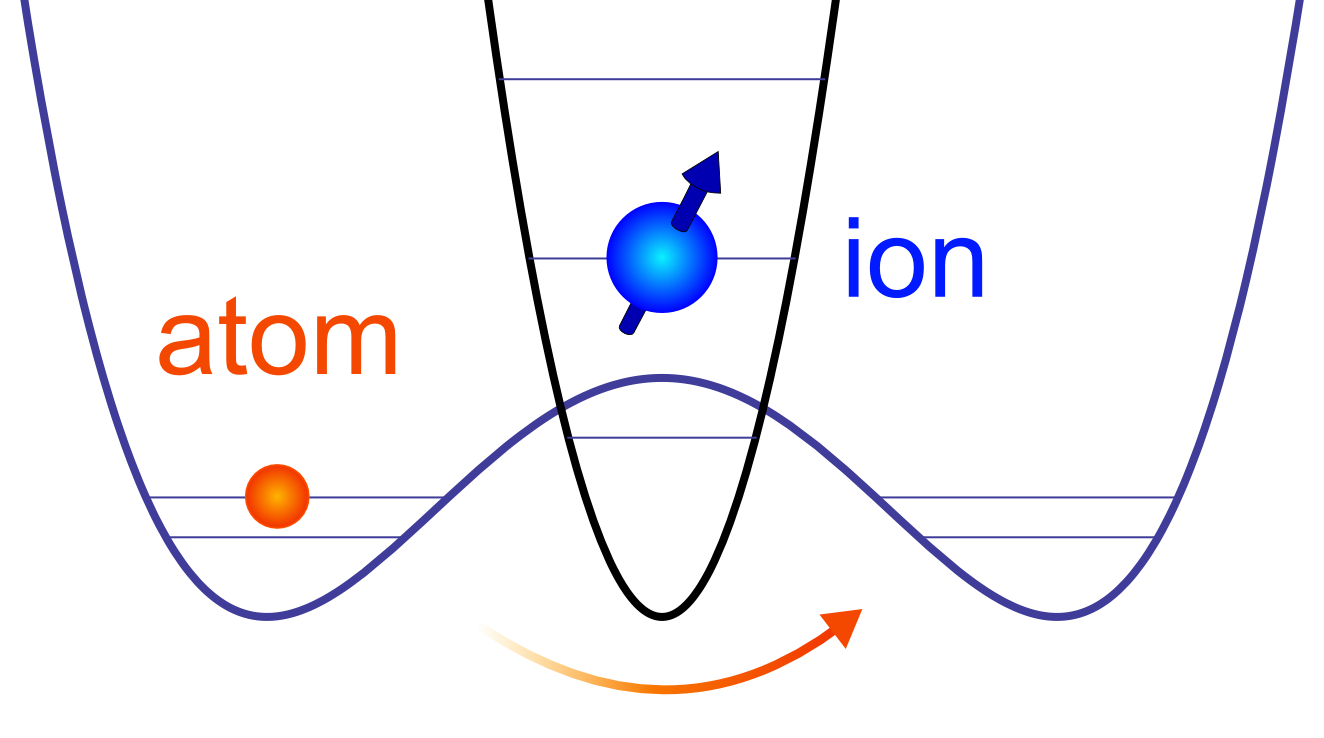} \caption{\label{FIG:junction}
(Color online) We consider a setup in which an atom is trapped in a double-well potential with a single ion trapped in its center. The quantised motional states of the ion and its internal spin state influence the atomic tunnelling rate between the wells. These degrees of freedom can in turn be manipulated with laser light or radio frequency fields.}
\end{figure}

In reference~\cite{Gerritsma:2012} we solved the atomic dynamics in the ion-controlled double-well system by assuming that the ion is pinned to the center of its trap. In this work we investigate how the dynamics of the ion changes the picture obtained in Ref.~\cite{Gerritsma:2012}. To this end, we consider a setup in which an atom is trapped in a double-well potential with a single ion trapped in its center as shown in Fig.~\ref{FIG:junction}. We will numerically solve the combined atom-ion dynamics in one dimension (1D) in terms of quantum defect theory (QDT) ~\cite{Idziaszek:2007,Idziaszek:2011}. This enables us to investigate the possibility of using non-classical states of the ion motion to control the tunnelling and the effect of imperfect ground-state cooling on the dynamics. We also address the role of micromotion - the fast oscillating motion of ions caused by a time-dependent Paul trap - on the tunnelling dynamics. This effect has been shown to significantly change the dynamics of the proposed atom-ion quantum gate~\cite{Doerk:2010}, as described in Ref.~\cite{Nguyen:2012}, leading to slower gates or requiring additional control pulses. Micromotion has also been shown to limit attainable temperatures for ions that are sympathetically cooled by atoms~\cite{Cetina:2012, Krych:2013}. Here we show that the micromotion causes difficulties in state-preparation and complicates the atomic tunnelling for large tunnelling rates, i.e. small inter-well separations. Additionally, we find that using an atom-ion combination with a large mass ratio, with the ion being the heavier particle, allows to overcome such difficulties. This conclusion is in line with a recent classical analysis studying attainable temperatures in atom-ion sympathetic cooling~\cite{Cetina:2012}.

The paper is structured as follows: In section~\ref{Sec_static} we will discuss the solutions to the atomic double-well in the presence of a static ion. In section~\ref{Sec_moving}, we will calculate how the situation changes when the ion dynamics are taken into account. We will show that the motional state of the ion is coupled to the atomic tunnelling rate such that the tunnelling can be controlled by engineering (non-classical) ionic states of motion. We will also analyze the effect of imperfect ion cooling. In section~\ref{Sec_mmotion} we will analyze the problem in the presence of a time-dependent trapping potential for the ion and the effect of micromotion. Then, we discuss possible experimental implementations and draw conclusions in Sec.~\ref{sec:expimp} and Sec.~\ref{sec:concl}, respectively. Finally, in the appendix we shall solve the double well problem in three dimensions (3D) and demonstrate that the 1D calculation results in the same physical picture.

\section{A static ion in 1D}\label{Sec_static}


The interaction between an atom and an ion is caused by an induced atomic dipole due to the electric field of the ion. At large distances it is given by:

\begin{equation}\label{eqPot}
\lim_{r\rightarrow \infty} V_{ia}(r)=-\frac{C_4}{r^4},
\end{equation}
\noindent with $C_4=e^2\alpha_p/2$. Here, $e$ is the charge of the ion and $\alpha_p$ is the static polarisability of the atom. It is useful to introduce the length scale $R^*=\sqrt{2\mu C_4/\hbar^2}$ and the energy scale $E^*=\hbar^2/[2\mu (R^*)^2]$ that characterize the atom-ion potential, with $\mu$ the reduced mass. For the atom-ion combinations studied in this work we have that $R^*=306$~nm and $E^*/h=935$~Hz for the atom/ion pair $^{87}$Rb/$^{171}$Yb$^+$~\cite{Steck:Rb}, and $R^*=75$~nm and $E^*/h=133$~kHz for the atom/ion pair $^{7}$Li/$^{171}$Yb$^+$~\cite{Miffre:2005}.

For $r\rightarrow 0$, Eq.~(\ref{eqPot}) does not describe the potential anymore as it becomes strongly repulsive. The exact form of the potential in this regime is generally not known well enough to solve the scattering dynamics, but as it has been shown in Ref.~\cite{Idziaszek:2007}, QDT can be employed to parametrize the potential at short-range (see Ref.~\cite{Idziaszek:2011} for more details). 

Although the potential~(\ref{eqPot}) is clearly spherically symmetric, we will for now limit ourselves to the one-dimensional case and we will denote the position of the ion (atom) as $z_i$ ($z_a$). We note that in 1D it turns out that the atom-ion interaction has the same mathematical expression as in 3D, that is $V_{ia}(z_i, z_a)=-C_4/(z_i-z_a)^4$~\cite{Idziaszek:2007}. We discuss the 3D scenario in the appendix and compare it to the 1D results. 

As an illustration of the atomic dynamics in a double-well system in the presence of an ion, we will first solve the system assuming that the ion is pinned to the center of its trap at $z_i=0$. Hence, the Hamiltonian of the atom is:

\beq
H_a=\frac{p_a^2}{2m_a}+V_{dw}(z_a) - \frac{C_4}{z_a^4},
\eeq

\noindent with $p_a$ the momentum of the atom, $m_a$ the atomic mass and $V_{dw}(z_a)$ the double-well potential. A convenient choice for $V_{dw}(z_a)$ is given by

\beq
\label{eq:Vdw}
V_{dw}(z_a)=\frac{b}{d^4}\left(z_a^2-d^2\right)^2.
\eeq

\noindent This potential has minima at $z_a=\pm d$ with local trapping frequencies $\omega_a=\sqrt{8b/(m_a d^2)}$ and inter-well barrier $b$. For the sake of simplicity and without loss of generality, we fix the local trapping frequency $\omega_a$ for each inter-well distance $2d$ by setting $b=\omega_a^2m_ad^2/8$. 

Our goal is to find a set of basis functions to expand the solution of $H_a$ onto, as this procedure is more efficient than solving the Schr\"odinger equation for each $d$ separately. To this end, we write the Hamiltonian as $H_a=H_a^{(0)}+H_a^{(1)}$ with:

\begin{eqnarray}\label{eq_Ha0}
H_a^{(0)}&=&-\frac{\hbar^2}{2m_a}\frac{\partial^2}{\partial z^2_a}+\frac{1}{2}m_a\omega^2_a z_a^2-\frac{C_4}{z^4_a},\\
\label{eq_Ha1}
H_a^{(1)}&=&\frac{1}{8}m_a\omega^2_a\left(d^2-2 z^2_a+\frac{z^4_a}{d^2}\right)-\frac{1}{2}m_a\omega^2_a z_a^2.
\end{eqnarray}
Note that the term $m_a\omega_a^2 z_a^2/2$ has been added to~(\ref{eq_Ha0}) and subtracted again in Eq.~(\ref{eq_Ha1}). This allows for finding a set of discrete basis states instead of the continuum that arises without this term for $E>0$. This is convenient as the final solutions of $H_a$ will form a discrete set as well~\cite{Gerritsma:2012}. 

To solve the Schr\"odinger equation for $H_a^{(0)}$ by means of QDT, we note that as $z_a\rightarrow 0$ the energy is dominated by the term $-C_4/z_a^4$. Therefore, we can neglect the other energies and solve the Schr\"odinger equation analytically. We obtain even and odd solutions given by~\cite{Idziaszek:2007}:

\begin{eqnarray}\label{eq_asymp_a}
\tilde{\psi}_e(z_a)&\propto&|z_a|\sin \left(\sqrt{\frac{m_a}{\mu}}\frac{R^*}{|z_a|}+\phi_e\right),\\
\label{eq_asymp_b}
\tilde{\psi}_o(z_a)&\propto& z_a \sin \left(\sqrt{\frac{m_a}{\mu}}\frac{R^*}{|z_a|}+\phi_o\right),
\end{eqnarray}

\noindent where $\phi_e$ and $\phi_o$ are the even and odd short-range phases, respectively. Since the short-range phases are not generally known experimentally and cannot be reliably obtained from ab-initio calculations, we choose a number of realistic values here, corresponding to scattering lengths in the range $-R^*$ to $R^*$. We also note that in a quasi 1D setup the values of the phases can be tuned by changing the confinement in the two remaining dimensions~\cite{Idziaszek:2007}.

In order to solve the Schr\"odinger equation for the Hamiltonian~(\ref{eq_Ha0}), we use Eqs.~(\ref{eq_asymp_a},\ref{eq_asymp_b}) as a boundary condition at some small value $z_{min}$ such that $C_4/z_{min}^4\gg~E_{max}$, with $E_{max}$ being the largest energy considered in the problem. A renormalized Numerov method~\cite{Johnson:1977} gives the solutions $\Phi^{(0)}_k(z_a)$ with energies $\mathcal{E}_k^{(0)}$ and quantum number $k$. To solve the dynamics in the double-well we now diagonalize the Hamiltonian with matrix elements:

\begin{eqnarray}
H_{kk'}&=&\!\left(\!E^{(0)}_{k}\!+\!\frac{m_a\omega_a^2d^2}{8}\!\right)\delta_{kk'}\!+\frac{m_a\omega_a^2}{4}\!\left(\frac{\!\mathcal{M}^{(4)}_{kk'}}{2d^2}\!-\!3 \mathcal{M}^{(2)}_{kk'}\!\right),\nonumber\\
\mathcal{M}^{(j)}_{kk'}&=&\!\int  \Phi^{(0)}_k(z_a)z_a^j \Phi^{(0)}_k(z_a) dz_a,\nonumber
\end{eqnarray}

\noindent with $j=2,4$ for a range of values $d$.


\begin{figure}
\includegraphics[width=8cm]{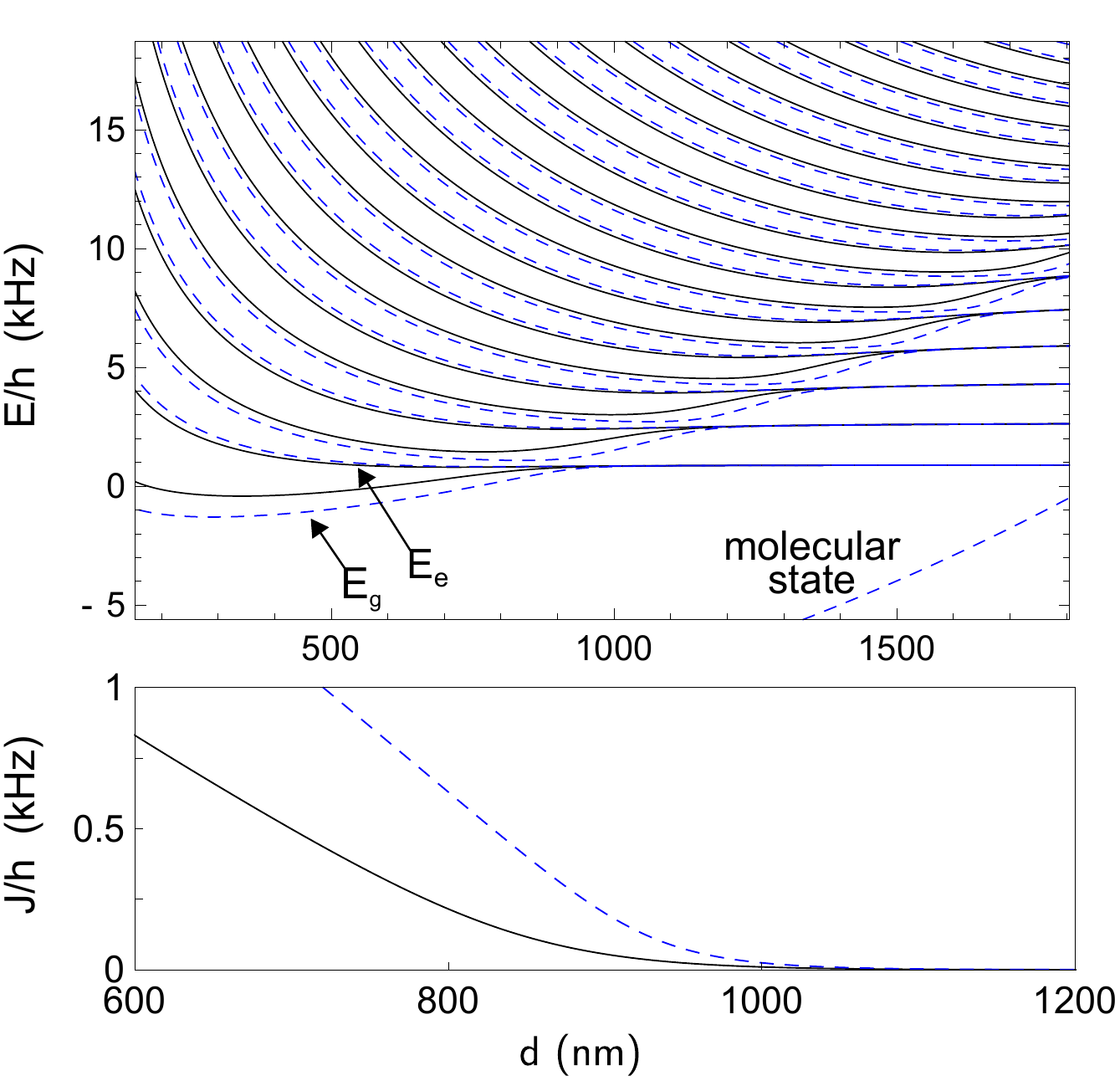} \caption{\label{FIG:atomonlyspec} (Color online)
Upper panel: Eigenenergy spectrum as a function of the inter-well separation $d$ for $^{87}$Rb and $^{171}$Yb$^+$ assuming $\phi_e=-\pi/4$, $\phi_o=\pi/4$ (black solid lines) and $\phi_e=-\pi/3$, $\phi_o=\pi/3$ (blue dashed lines). For large well separations, the spectrum resembles that of two independent harmonic oscillators of the same trap frequency $\omega_a=2\pi\times$1.8~kHz. As the wells approach, each level splits in two owing to the energy difference of even and odd wavefunctions.  In the right corner a molecular state can also be seen. The occuring tunnelling rate is given by the energy difference and is plotted in the lower panel.
}
\end{figure}

Now, as in our previous study~\cite{Gerritsma:2012}, we consider $^{87}$Rb and $^{171}$Yb$^+$ as an example system and we assume that no spin-changing collisions can occur such that our single channel description is accurate. Setting $\omega_a=2\pi\times$~1.8~kHz we obtain the spectrum shown in Fig.~\ref{FIG:atomonlyspec}. For large inter-well separations $d$ the eigenenergies resemble the equidistant level structure of the harmonic oscillator as the wells are uncoupled. While decreasing the inter-well distance, the energy levels split in two as there is an energy difference between states of even and odd symmetry. For a single atom and small energy splitting, this energy difference corresponds to the inter-well tunnelling rate in a two-mode picture~\cite{Millburn:1997}. Since the barrier height $b$ decreases with distance, the higher excited atomic states exhibit level splitting at larger separation $d$ than low energy states. The eigenenergies that asymptotically connect to the harmonic oscillator ground states correspond to superpositions of approximately localized wavepackets: $\Phi_{e,g}(z_a)=(\Phi_L(z_a)\pm\Phi_R(z_a))/\sqrt{2}$ in the left (L) and right (R) well. These states are labeled with the subscripts $g$ and $e$ and have energies $E_{e,g}$  with $E_e \geq E_g$. At $d=$~900~nm we find that the ground state degeneracy is lifted corresponding to a tunnelling rate of~$J/h=(E_e-E_g)/h=$~56~Hz when $\phi_e=-\pi/4$, $\phi_o=\pi/4$ (black lines) and ~$J/h=$~202~Hz when $\phi_e=-\pi/3$, $\phi_o=\pi/3$ (dashed blue lines). Since the short-range phases depend on the relative spin orientations of the atom and ion, the above calculation demonstrates a similar ion spin-dependence as in Ref.~\cite{Gerritsma:2012}. Finally, we note that the spectrum was obtained by taking 106 basis states into account.

In appendix~\ref{app_3D_M} we perform a 3D analysis of the double-well for a static ion that we compare to the 1D model of the present section. We show that the double-well problem in 1D has a strict analogy to the 3D scenario, and therefore it provides a satisfactory physical picture of the system. Given this, we extend our analysis to the case of a moving ion. A 3D study of such a system would be indeed rather difficult to treat numerically and nonetheless it would not provide further insight into the problem.

\section{A moving ion}\label{Sec_moving}

Now we will see how the picture is altered when we allow the ion to move. As outlined above, we will restrict our attention to the 1D scenario. In this case the Hamiltonian is given by:

\beq\label{H_dw_moving}
H=\frac{p_i^2}{2m_i}+\frac{p_a^2}{2m_a}+\frac{1}{2}m_i\omega_i^2z_i^2+V_{dw}(z_a)-\frac{C_4}{(z_i-z_a)^4}.
\eeq

\noindent Here, $\omega_i$ denotes the ion trap frequency, $m_i$ the mass of the ion, and $p_i$ its momentum. To find the eigenstates and energies of this Hamiltonian, we write the Hamiltonian in terms of relative and center-of-mass coordinates $r=z_i-z_a$, $R=(m_iz_i+m_az_a)/M$ with $M=m_i+m_a$ the total mass, as $H^{(d)}=H_R^{(0)}+H^{(0)}_r+H^{(1)}$:

\begin{eqnarray}
H_R^{(0)}&=&-\frac{\hbar^2}{2M}\frac{\partial^2}{\partial R^2}+\frac{1}{2}M\omega_R^2R^2\label{eq_H0R}\\
H^{(0)}_r&=&-\frac{\hbar^2}{2\mu}\frac{\partial^2}{\partial r^2}+\frac{1}{2}\mu\omega_r^2r^2-\frac{C_4}{r^4}\label{eq_H0r}\\
H^{(1)}&=&\mu(\omega_i^2-\omega_a^2)Rr+V_{dw}(R,r) \nonumber\\
&-&\frac{\mu\omega_a^2}{2}\left(\frac{m_a}{\mu}R^2+\frac{\mu}{m_a}r^2-2Rr\right).\label{eq_H1}
\end{eqnarray}

\noindent  Equation~(\ref{eq_H0R}) denotes the Hamiltonian of a harmonic oscillator with trap frequency $\omega_R^2=(m_i\omega_i^2+m_a\omega_a^2)/M$. It has eigenstates $f_n(R)$ and energies $E^{(0)}_n=\hbar\omega_R(n+1/2)$. Equation~(\ref{eq_H0r}) is similar to equation~(\ref{eq_Ha0}) except for some pre-factors, in particular $\omega_r^2=(m_i\omega_a^2+m_a\omega_i^2)/M$. As explained in the previous section, we have determined the eigenfunctions $\Phi^{(0)}_k(r)$ and eigenenergies $\mathcal{E}^{(0)}_k$ of such Hamiltonian by means of QDT. To find the eigenstates and eigenenergies to the full Hamiltonian we expand onto the basis $|f_n,\Phi^{(0)}_k\rangle$. We denote the solutions for each inter-well separation $d$ as $\psi_l^{(d)}(R,r)=\sum_{nk}c^{(d)}_{lnk} f_n(R)\Phi^{(0)}_k(r)$ with energies $E_l^{(d)}$ and coefficients $\mathbf{c}^d_{nk}$, and where an overall quantum number $l$ labels each solution.

\subsection{Examples}

\begin{figure}
\includegraphics[width=8.5cm]{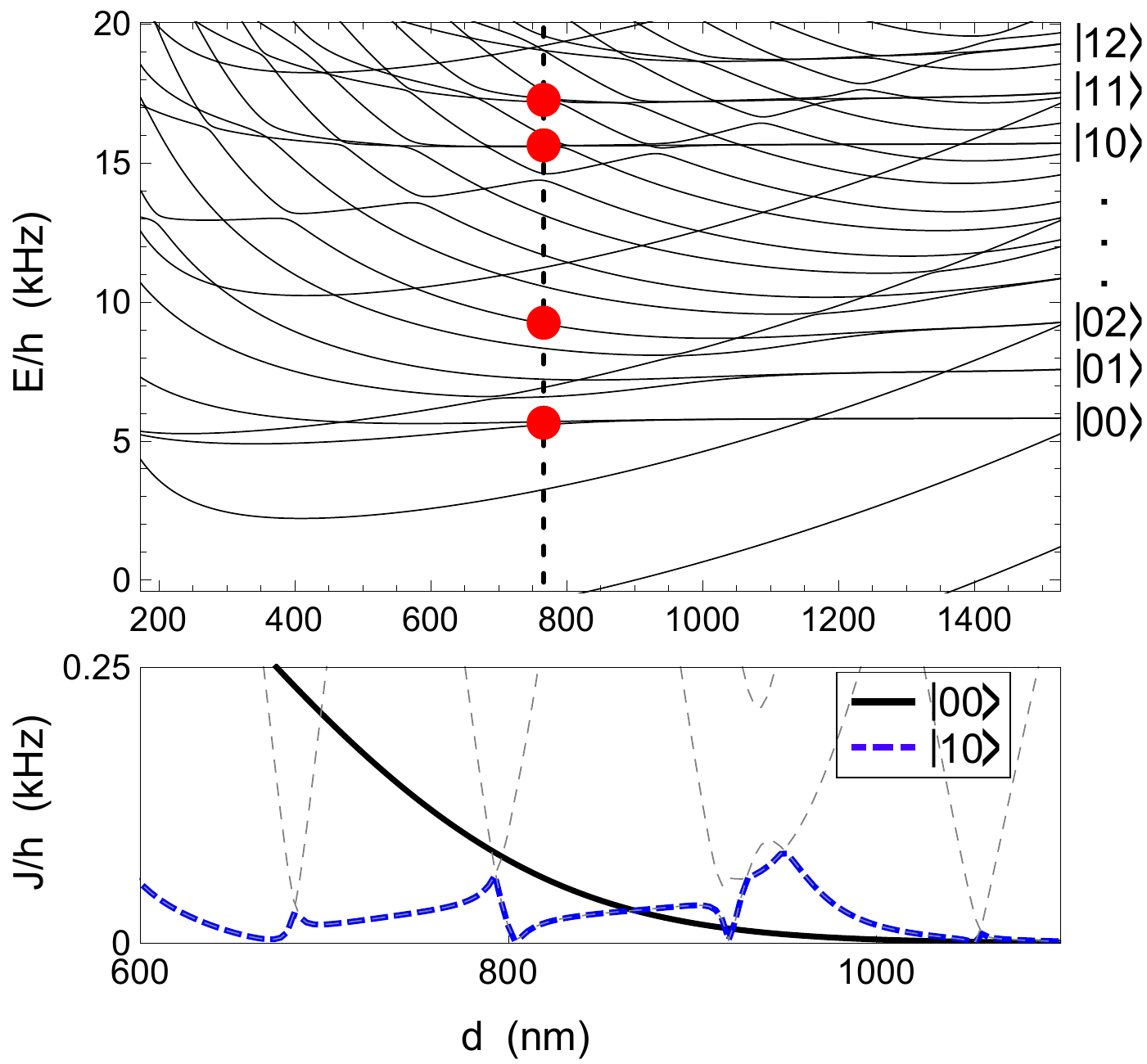}
\caption{\label{FIG:ionmovingspecwithandwithoutIA}
(Color online)  Upper panel: Spectrum of a double-well system where the ion is allowed to move for $^{87}$Rb and $^{171}$Yb$^+$ with $\omega_a=2\pi \times 1.8\,$kHz, $\omega_i=2\pi \times 9.9\,$kHz, $\phi_e=-\pi/4$, and $\phi_o=\pi/4$.
On the right, we give some of the asymptotic quantum numbers $|n_i,n_a\rangle$ for large well separation. The red dots correspond to the wavefunctions plotted in Fig.~\ref{fig:wavs2D_q2p56_Rb_Yb_-pi4_pi4_n0m0}. In the lower panel, the level splitting is plotted for the case where the ion is in the ground state (thick black line) and for the case where it is in the first excited state (dashed lines). When avoided crossings occur, the energy level separation becomes dependent on whether the crossing is traversed diabatically or adiabatically. The smallest $J$ is plotted using a thicker dashed blue line. Close to the avoided crossings, the two mode description is no longer accurate and $J$ loses its interpretation as the tunnelling rate.}
\end{figure}

As an example of the resulting spectrum we show the situation for $^{87}$Rb and $^{171}$Yb$^+$ assuming $\phi_e=-\pi/4$, $\phi_o=\pi/4$, $\omega_a=2\pi \times 1.8\,$kHz and $\omega_i=2\pi \times 9.9\,$kHz in Fig.~\ref{FIG:ionmovingspecwithandwithoutIA}. This spectrum was obtained by expanding the solutions onto 6417 basis states (93 states in the relative coordinate and 69 states in the center-of-mass coordinate). In comparison to Fig.~\ref{FIG:atomonlyspec} we see that more energy levels appear, both in the form of molecular states and trap states. When the wells are far apart, the states reduce to the harmonic oscillator states for the atom and ion. We can identify the asymptotic atom-ion Fock states $|n_{i},n_{a}\rangle$ in Fig.~\ref{FIG:ionmovingspecwithandwithoutIA} for large $d$. For intermediate $d$, the states are perturbed by the atom-ion interaction. 

Focusing on the states that connect to the harmonic oscillator ground states $|00\rangle$, we see the same behaviour as for the case where the ion was static: The initial degeneracy for large inter-well separations is lifted as the wells come closer and a tunnelling rate can be identified.  Molecular states cross the ground states at a few points, but the avoided crossings are very small. Therefore, the static ion approximation was justified when considering the ground states.

The state connecting to $|10\rangle$ shows a similar behaviour, but more and larger avoided crossings appear. Close to these crossings, a simple two-mode picture breaks down and the tunnelling rate $J$ loses its meaning. In an experiment, the inter-well separation $d$ would typically be reduced dynamically to initiate tunnelling. The resulting tunneling rate then depends on whether the crossing is traversed diabatically or not. For most values of $d$, the energy splitting is different than for the case when the ion is in the ground state. For $d=775$~nm, for instance, we get a tunnelling rate of $J/h=101$~Hz when the ion is in the ground state, but only $J/h=37$~Hz when the ion is in the first Fock state. Therefore, the atomic tunnelling rate depends on the motional state of the ion. 

\begin{figure}
\includegraphics[width=8.5cm]{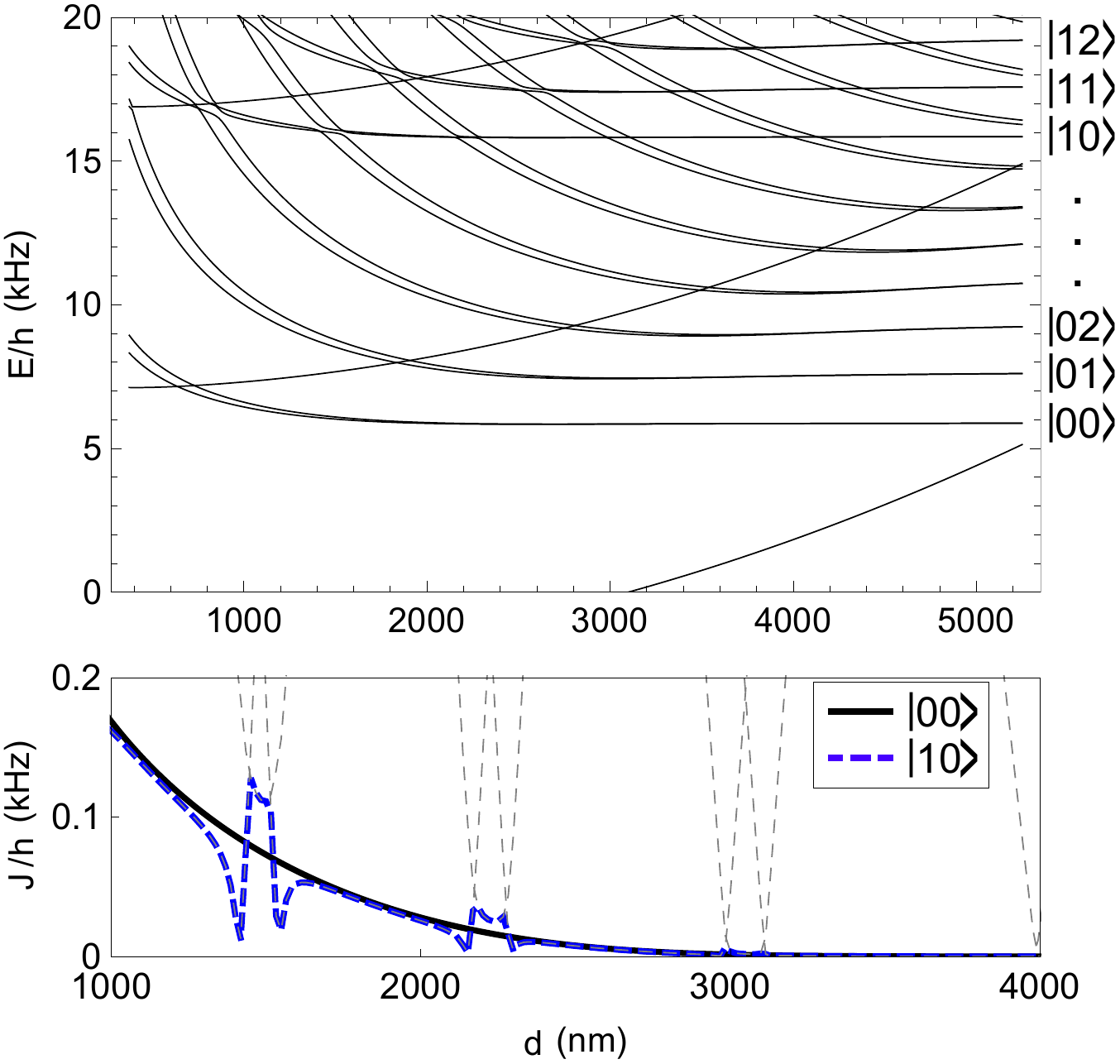}
\caption{\label{FIG:ionmovingspecYbLi}
(Color online)  Upper panel: Spectrum of a double-well system where the ion is allowed to move for $^{7}$Li and $^{171}$Yb$^+$ with $\omega_a=2\pi \times 1.8\,$kHz, $\omega_i=2\pi \times 9.9\,$kHz, $\phi_e=-\pi/4$, and $\phi_o=\pi/4$, i.e. the same as in Fig.~\ref{FIG:ionmovingspecwithandwithoutIA}. Note that tunnelling occurs for larger distances than for $^{87}$Rb as the $^7$Li wavepacket is larger for the same trapping frequency. On the right, we give the asymptotic quantum numbers $|n_i,n_a\rangle$ for large well separation. The lower panel shows the energy level separation $J$ for the ion in the ground state (black solids line) and the first excited state (dashed blue). The avoided crossings are also indicated by the thin grey lines. Clearly, in this case the tunnelling rate is very similar for the first excited and ground states when no avoided crossings are near. 
}
\end{figure}

As a second example, we also plot the case for $^{7}$Li and $^{171}$Yb$^+$ assuming $\phi_e=-\pi/4$, $\phi_o=\pi/4$, $\omega_a=2\pi\times~1.8\,$~kHz and $\omega_i=2\pi\times9.9\,$~kHz (see Fig.~\ref{FIG:ionmovingspecYbLi}). We see that the tunnelling occurs for larger inter-well separations for these species as the atomic wavepacket is larger because of its lower mass. Furthermore, the tunnelling rate shows less dependency on the motional state of the ion as long as there are no avoided crossings nearby. This is a consequence of the large mass ratio between the atom and ion that separates atomic and ionic dynamics in a similar fashion as in the Born-Oppenheimer approximation.  We note that in this case 2700 states were taken into account.

\begin{figure}
\includegraphics[width=7.5cm]{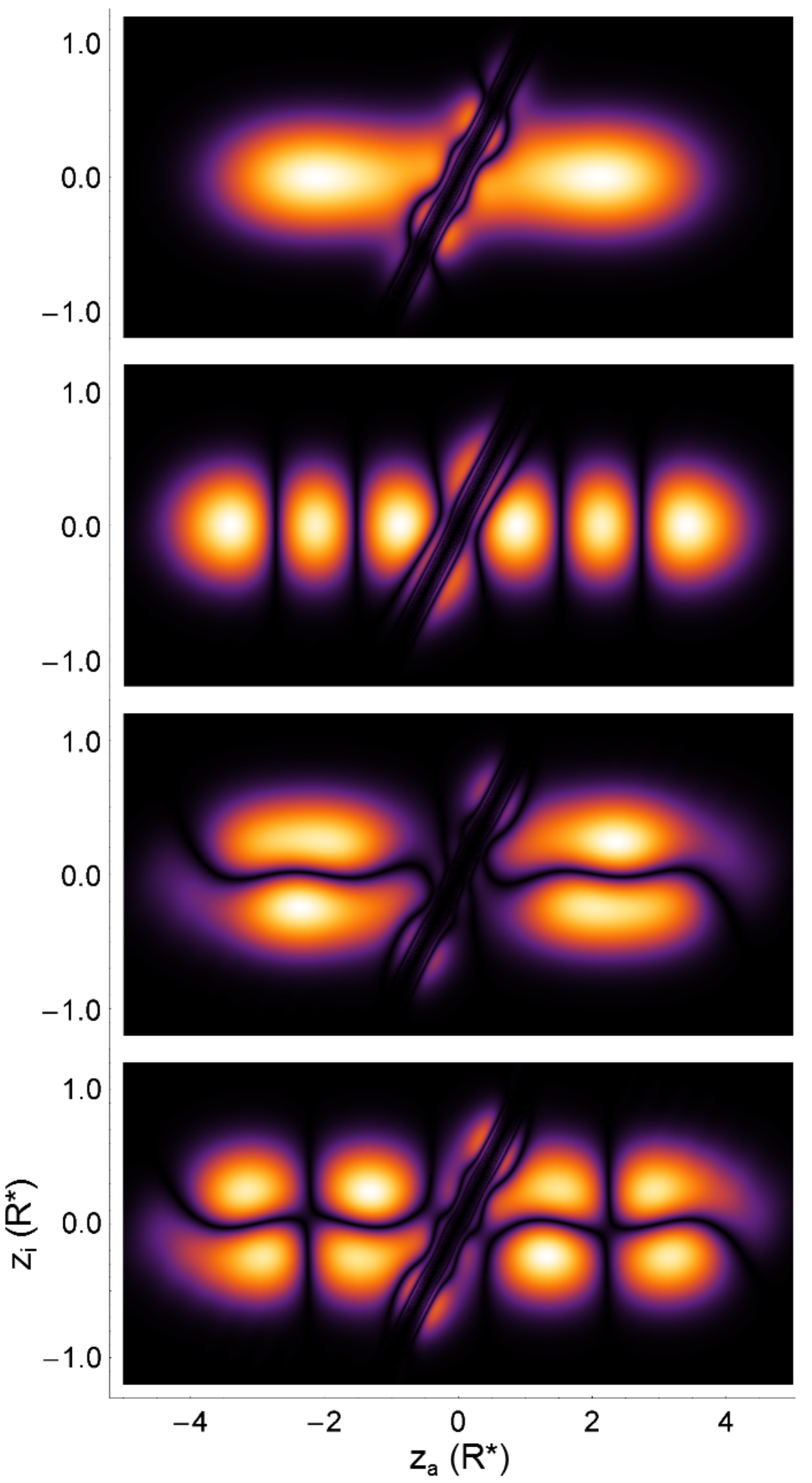}
\caption{\label{fig:wavs2D_q2p56_Rb_Yb_-pi4_pi4_n0m0}
(Color online)  Wavefunctions $|\psi_l^{(d)}(z_i,z_a)|$ of the ion-atom states for an inter-well separation of $d=2.5 R^*$=765~nm for $^{87}$Rb and $^{171}$Yb$^+$ with $\phi_e=-\pi/4$, $\phi_o=\pi/4$, $\omega_a=2\pi\times 1.8\,$kHz, and $\omega_i=2\pi\times 9.9\,$kHz, corresponding to the red dots in Fig.~\ref{FIG:ionmovingspecwithandwithoutIA}. We used units of $R^*$ on the axes as these are the natural units for the atom-ion system. The states asymptotically correspond to the Fock states $|n_{i},n_{a}\rangle= |00\rangle$, $|02\rangle$, $|10\rangle$ and $|11\rangle$ for large well separation (with the atom occupying both wells). The top wavefunction is the double-well ground state $|\Phi_{g}\rangle$. The corresponding quantum numbers can still be recognised by counting the nodes in the direction of the ionic and atomic coordinate. In the vicinity of the line $z_a=z_i$ the atom-ion interaction potential dominates and the wavefunctions take similar forms as in Eqs.~(\ref{eq_asymp_a},\ref{eq_asymp_b}). The fast oscillations around this region are not completely resolved in the density plot. 
}
\end{figure}

In Fig.~\ref{fig:wavs2D_q2p56_Rb_Yb_-pi4_pi4_n0m0} we plot a few probability distributions for the eigenstates corresponding to the red dots in Fig.~\ref{FIG:ionmovingspecwithandwithoutIA}. We see that for $z_a\sim z_i$, the atom-ion interaction dominates, but for larger atom-ion distances the probability distributions resemble those of the harmonic oscillator in each well and the presence of the ion only slightly perturbs the double-well system

\begin{figure}
\includegraphics[width=8cm]{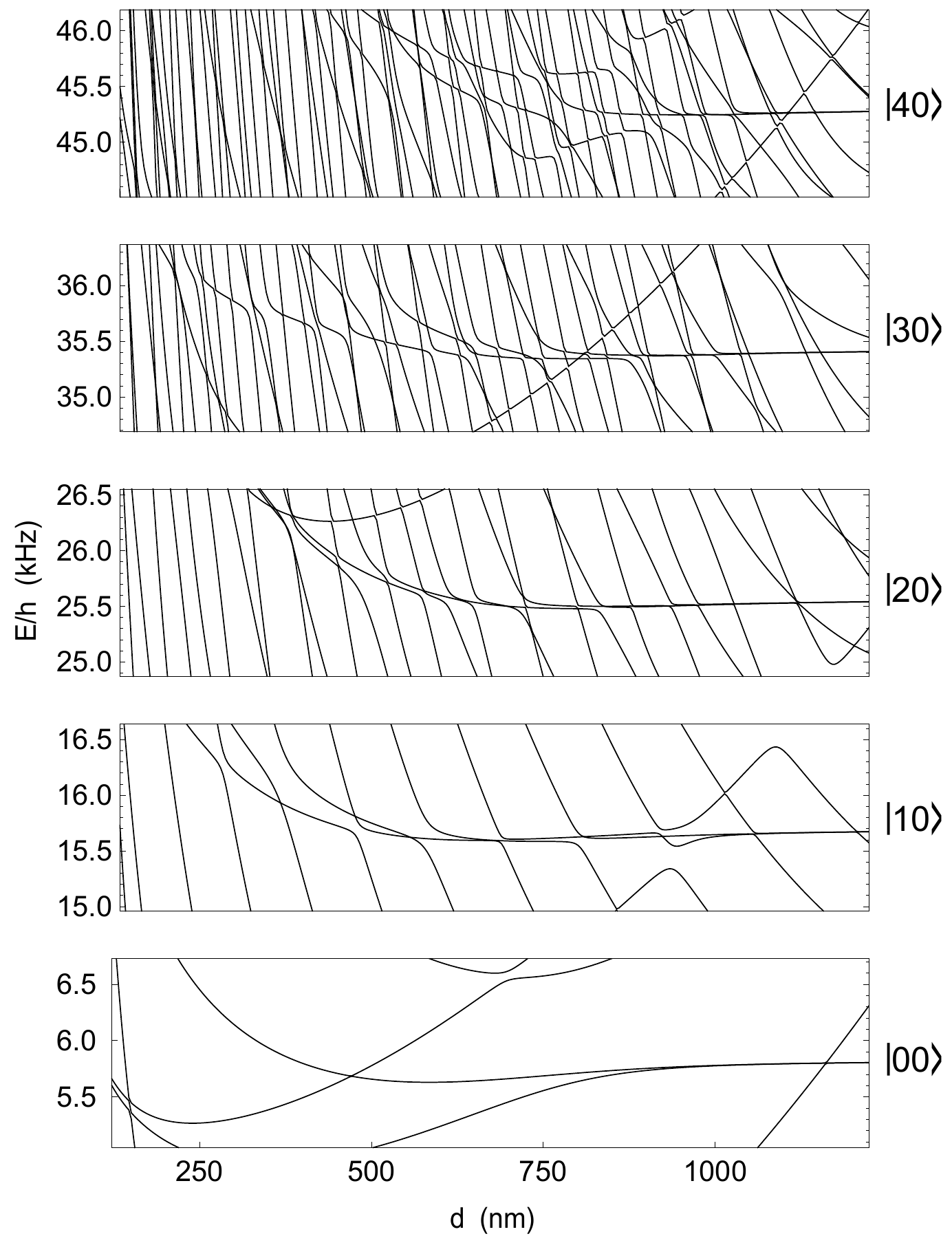}
\caption{Enlarged view of the five first Fock states of the ion in the spectrum of a double-well with moving ion for $^{87}$Rb and $^{171}$Yb$^+$ with $\phi_e=-\pi/4$, $\phi_o=\pi/4$, $\omega_a=2\pi\times 1.8\,$kHz and $\omega_i=2\pi\times 9.9\,$kHz. }
\label{fig:spec_RbYb_gam5p5_-pi4_pi4_alpha10_n_all}
\end{figure}

\subsection{Controlling tunnelling with non-classical states of ion motion}

An interesting application using the dependence of the tunnelling rate on the motional state of the ion is the generation of entanglement between the atom and the ion. In Ref.~\cite{Gerritsma:2012} the state dependent tunnelling was proposed via the state-dependence of the short-range phase. Alternatively, we can create non-classical states of ion motion, such as Fock states, to engineer motional state-dependent tunnelling. The Fock state can in turn be entangled with the spin of the ion. Such quantum states are commonly engineered and analyzed in ion trap experiments~\cite{Leibfried:1996,Ziesel:2013}.

\begin{figure}
\includegraphics{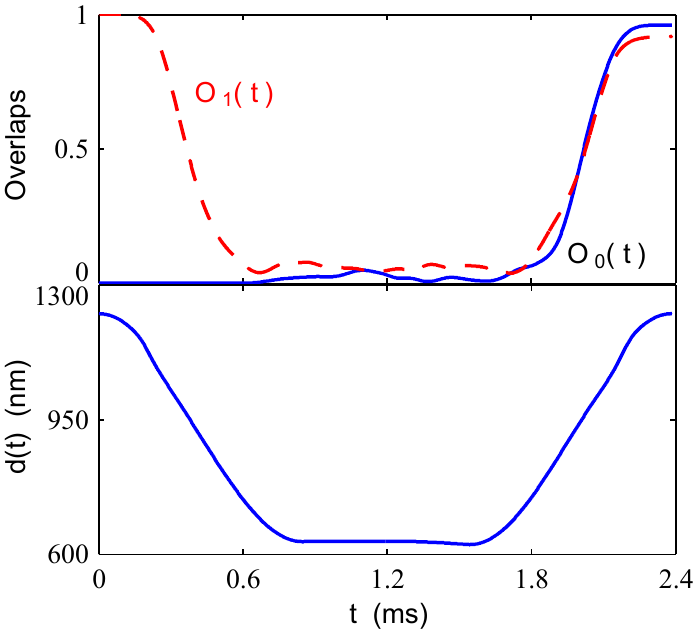}
\caption{\label{fig:entanglement}
(Color online) Optimisation of the entanglement generation protocol as discussed in the main text. Upper panel: Overlaps $O_{0,1}(t)$ for the optimised dynamics. Lower panel: Optimal inter-well separation. The optimisation has been performed for the atom-ion pair $^{87}$Rb/$^{171}$Yb$^+$ with $\phi_e=-\pi/4$ and $\phi_o=\pi/4$. 
}
\end{figure}

From the spectrum displayed in Fig~\ref{FIG:ionmovingspecwithandwithoutIA}, we see that a good strategy to engineer atom-ion entanglement consists of the following steps: 

\begin{enumerate}
\item We first prepare the ion in an equal superposition of the ground and first excited state of its trap, while its internal state is prepared in a specific hyperfine level.
\item Subsequently, we prepare the atom, for instance, in the ground state of the left well and far away from the ion such that the tunnelling is negligible. 
\item Then we reduce the inter-well separation $d$ dynamically until tunnelling occurs.
\item Afterwards, we wait until the atom has tunnelled to one side depending on the ion Fock state, that is, to the right well if the ion Fock state is $\ket{0}$ and back to the left well if the ion Fock state is $\ket{1}$ (see Fig.~\ref{FIG:ionmovingspecwithandwithoutIA} first and third red dots, from below). 
\end{enumerate}
For the ion in the ground state, $n_i=0$, we see that the sequence is very similar to the one presented in Ref.~\cite{Gerritsma:2012} (see Fig.~\ref{FIG:ionmovingspecwithandwithoutIA}). For higher Fock states, however, we see an increasing amount of avoided crossings. To pass these crossings diabatically during our sequence, an accurate control of the inter-well distance $d(t)$ is required. We see that for $n_i=1$ the number of crossings is still quite limited and we take this as an example. We have performed numerical simulations of the dynamics and we have optimised the process, by means of the chopped random-basis algorithm~\cite{Caneva:2011}, in order to maximise the entanglement. To this aim, we have defined the following overlaps:

\begin{align}
O_{0}(t)=\vert\langle\psi_{R,0}^{(0)}\vert\psi_{L,0}(t)\rangle\vert^2,\nonumber\\
O_{1}(t)=\vert\langle\psi_{L,1}^{(0)}\vert\psi_{L,1}(t)\rangle\vert^2.
\end{align}
Here, $\ket{\psi_{R,0}^{(0)}}$ is the wave function of the atom in the ground state of the right well and the ion in the ground state of its harmonic trap, whereas $\ket{\psi_{L,1}^{(0)}}$ is the wave function of the atom in the ground state of the left well and the ion in the first excited state of its harmonic trap. The time evolved states are obtained from $\vert\psi_{L,0}(t)\rangle =  U(t) \ket{\psi_{L,0}^{(0)}}$ and $\vert\psi_{L,1}(t)\rangle =  U(t) \ket{\psi_{L,1}^{(0)}}$, where $ U(t)$ is the time evolution operator generated by the Hamiltonian~(\ref{H_dw_moving}). In the optimization procedure we minimised the overlap infidelity, $2- O_0(T)-O_1(T)$, at the final time $T$, that is, the time needed to perform the sequence 3.-4. outlined above, by optimally controlling the inter-well separation $d(t)$. In Fig.~\ref{fig:entanglement} we show the overlaps $O_{0,1}(t)$ for the optimal inter-well separation $d(t)$. We obtained $O_0(T) \simeq 0.96$ and $O_1(T)\simeq 0.92$ with $T\simeq 2.38$~ms for the atom-ion pair $^{87}$Rb/$^{171}$Yb$^+$. This result shows that we can produce atom-ion entanglement in a very short time by means of the ionic motional state. We note, however, that the overlaps are not perfect. This might be achieved by controlling, for instance, the atom-ion interaction via Feshbach resonances, or by controlling additionally the height of the barrier. Our goal here, however, is to show that such entanglement generation is in principle possible. A more detailed analysis by means of optimal control theory would require perfect knowledge of the trapping potentials, and not just the analytically convenient form given by Eq.~(\ref{eq:Vdw}). 

\subsection{Imperfect ground state cooling}

Up until now, we have assumed that both the atom and the ion are prepared in a pure state by ground state cooling (for instance via resolved sideband cooling) followed by coherent state manipulation. To see what the effect of imperfect ground state cooling of the ion would be, we have plotted in Fig.~\ref{fig:spec_RbYb_gam5p5_-pi4_pi4_alpha10_n_all} the spectra for the first 5 states that asymptotically correspond to the first 5 Fock states of the ion. When the ion is in a thermal state, each of these Fock states contributes to the tunnelling rate in a incoherent manner. We see that for $n_{i}=0,..,3$ the energy levels display degeneracy in $|\Phi_{e,g}\rangle_{n_i}$ for large $d$, in analogy to the situation for the ground states discussed above. As the inter-well separation is decreased these degeneracies are lifted and tunnelling occurs. Because of the coupling between the atomic tunnelling and the motion of the ion, each ionic state corresponds to a different tunnelling rate. Unless only few Fock states are occupied in the thermal state or the tunnelling rates are very similar for all Fock states, the coherence in the atomic state will be destroyed after a while. 

For the states of higher energy $n_{i}\geq 3$ we see that many avoided crossings start appearing that cannot be ignored anymore for small $d$. In this situation, the simple two mode picture cannot be employed anymore. 

\begin{figure}
\includegraphics{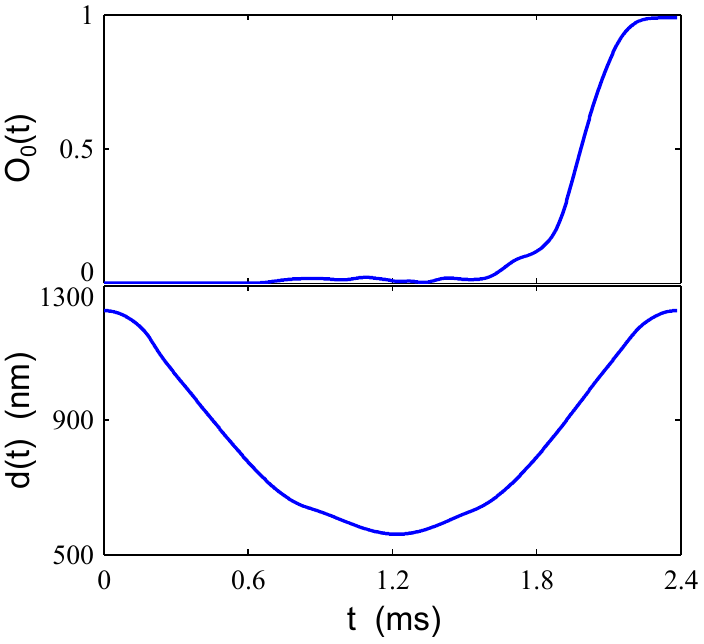}
\caption{\label{fig:tunnelling}
(Color online) Optimisation of the tunnelling dynamics when the ion is prepared in the ground state. Upper panel: Overlap $O_{0}(t)$ for the optimised dynamics. Lower panel: Optimal inter-well separation. The optimisation has been performed for the atom-ion pair $^{87}$Rb/$^{171}$Yb$^+$ with $\phi_e=-\pi/4$ and $\phi_o=\pi/4$. The final overlap fidelity is $O_{0}(T)\simeq 0.99$. 
}
\end{figure}

As an example, we have focused our attention to the tunnelling only, in particular when the ion is (ideally) prepared in the ground state. To this end, we performed first an optimisation of this dynamics. The corresponding result is shown in Fig.~\ref{fig:tunnelling}. In this case we were able to achieve an overlap fidelity of about $O_0(T) \simeq 0.99$ in $T\simeq 2.38$ ms. 
Given this result, we have investigated the impact of finite temperature on the tunnelling dynamics. To begin with, we have defined the initial density matrix of the ion as

\begin{align}
\rho_i(\mathcal{T}) = \sum_{n_i}p_{n_i}(\mathcal{T}) \vert n_i\rangle\langle n_i\vert
\end{align}
where $ \vert n_i\rangle$ are the harmonic oscillator eigenstates of the ionic harmonic trap, $p_{n_i}(\mathcal{T})$ are the occupation probabilities which are calculated by assuming a thermal distribution corresponding to temperature $\mathcal{T}$ in the canonical ensemble. The overlap fidelity at time $t=T$ for a given temperature is given by 

\begin{align}
\label{eq:FT}
F(\mathcal{T}) = \sum_{n_i}p_{n_i}(\mathcal{T}) \vert\langle\psi_{R,n_i}\vert\psi_{L,n_i}(t)\rangle\vert^2.
\end{align}
Here, $\vert\psi_{L,n_i}(t)\rangle$ is the time evolved state assuming that the initial state is the ground state of the left well for the atom and the $n$-th state of the harmonic trap for the ion. 

\begin{figure}
\includegraphics{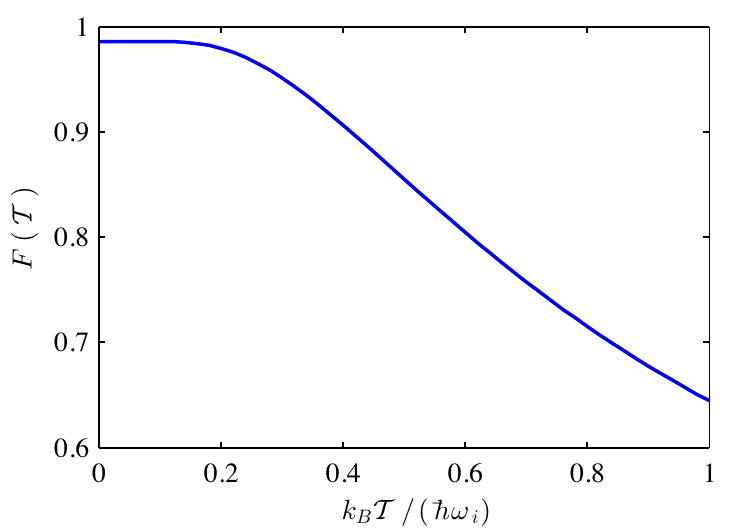}
\caption{\label{fig:temperature}
(Color online) Fidelity $F(\mathcal{T})$ of the tunnelling dynamics for the atom-ion pair $^{87}$Rb/$^{171}$Yb$^+$, whose overlap fidelity when the ion is prepared in the ground state of the trap and the corresponding optimal inter-well separation are displayed in Fig.~\ref{fig:tunnelling}. 
}
\end{figure}

It is interesting to see values of the fidelity for temperatures up to $k_B \mathcal{T}\approx \hbar\omega_i$ ($k_B$ is the Boltzmann constant). Let us define $\gamma = \exp(-\hbar\omega_i/k_B\mathcal{T})$ and neglect terms of $o(\gamma^5)$ in Eq.~(\ref{eq:FT}), since for higher energies the motional states of the atom and the ion are not anymore separable. Such temperature dependence of the fidelity is displayed in Fig.~\ref{fig:temperature}, for which we used the optimal inter-well separation shown in the lower panel of Fig.~\ref{fig:tunnelling}. As it is shown, the fidelity drops rather quickly, since the overlaps at time $T$ for the ionic states $n_i=1,2,3,4$ are significantly reduced. This shows that the tunnelling rates are different for those Fock states, and therefore the coherence in the atomic state is destroyed rapidly. 

A strategy to reduce the impact of finite temperature could be to perform an optimisation in which the optimal inter-well separation $d(t)$ is engineered in such a way that $F(\mathcal{T})$ is maximised within a given temperature range. This decoherent dynamics will be studied in the future more systematically via a quantum open system approach. 

\section{Micromotion}\label{Sec_mmotion}

So far, we have assumed that the ion is trapped in a time-independent harmonic trap. In most atom-ion experiments, however, the ions are trapped in a Paul trap that is based on an oscillating electric field. This causes a rapid oscillating motion of the ion called micromotion.

The Hamiltonian describing this situation in 1D is given by:

\begin{equation}\label{H_ion}
H_{ion}(t)=\frac{p_i^2}{2m_i}+\frac{1}{8}m_i\Omega^2z_i^2\left[a+2q\cos(\Omega t)\right].
\end{equation} 

\noindent Here, the parameters $a$ and $q$ depend on the trapping potential, the ion mass and its charge. With properly chosen $a$ and $q$, an effective time-independent trapping potential can be derived via the so-called secular approximation. In this approximation the ion motion reduces to a combination of the slow (secular) motion, that is, of a particle in a time independent trap, and a rapid micromotion of small amplitude. In most experiments the following inequalities hold: $a\ll q < 0.91$, where $q$ lies typically in the range 0.1-0.4. 

In order to evaluate the impact of the time-dependent ion trap on the double-well system, we replace the ion trap term (i.e., the kinetic and trapping potential energies) in Eq.~(\ref{H_dw_moving}) by the time dependent version of Eq.~(\ref{H_ion}). Nguyen {\it et al.}~\cite{Nguyen:2012} have analysed a  situation in which the atom was trapped in a harmonic trap. Here we will use the same  approach, but for a double-well potential. We perform the transformation of Cook and Shankland~\cite{Cook:1985,Nguyen:2012} to obtain a Hamiltonian that is comprised of a time-independent and a time-dependent term that we label with $mm$ (micromotion) $H_{tot}=H^{(d)}+H_{mm}(t)$. Note that $H^{(d)}$ is the Hamiltonian of the double well without micromotion, which is defined in Eqs.~(\ref{eq_H0R},\ref{eq_H0r},\ref{eq_H1}). More details on this procedure can be found in the Appendix~\ref{App_mm}. The micromotion Hamiltonian in the center-of-mass and relative coordinates is given by:

\begin{eqnarray}
H_{mm}(t)&=&-m_ig^2\omega_i^2\left(R^2+\frac{\mu^2}{m_i^2}r^2+\frac{2\mu}{m_i}rR\right)\cos 2\Omega t\nonumber\\
&-&g\omega_i\left(\{R,p\}+\frac{\mu}{m_i}\{r,p\}+\frac{m_i}{M}\{R,P\}\right.\nonumber\\
&+&\left.\frac{\mu}{M}\{r,P\}\right) \sin \Omega t.
\end{eqnarray}

\noindent Here $g=[2(1+2a/q^2)]^{-1/2}$ and $\{.,.\}$ denotes the anti-commutator. The relative and center-of-mass momenta are denoted by $p$ and $P$, respectively.  

Following again Ref.~\cite{Nguyen:2012}, we use Floquet theory to obtain the energies and eigenstates in terms of the unperturbed eigenstates of the Hamiltonian $H^{(d)}$. It turns out that we have to diagonalise the Hamiltonian

\beq
H_F=H^{(d)}+H_{mm}(t)-i\hbar\frac{\partial}{\partial t}.
\eeq

\noindent We use the unperturbed Floquet eigenstates  $\vert u_{jl}\rangle=e^{ij\Omega t}|\psi_l^{(d)}(R,r)\rangle$ of $H_F-H_{mm}(t)$ as our basis with Floquet energies $\epsilon_{jl}=E_l+j\hbar\Omega$. Here the integer $j$ denotes the class of the Floquet state, whereas the quantum number $l$ denotes the solution to the double-well problem without micromotion. Then, we introduce the generalised matrix elements:

\beq
\label{eq:genmatrxel}
 \langle \langle u^*_{j'l'}|H_{mm}(t)|u_{jl} \rangle
 \rangle=\frac{1}{T}\int_0^T dt \langle u^*_{j'l'}|H_{mm}(t)|u_{jl}\rangle,
\eeq

\noindent where now $T$ indicates the period of the micromotion. To gain further insight into the micromotion effect we write the matrix elements as follows~\cite{Nguyen:2012} (see Appendix~\ref{App_mm}):

\begin{widetext}
\begin{eqnarray}
 \langle u^*_{j'l'}|H_{mm}(t)|u_{jl} \rangle
 &=&\langle \psi^{(d)}_{l'} |\left[V_1\cos (2\Omega t)+ V_2 \sin (\Omega t)  \right]e^{i(j-j')\Omega t}|\psi^{(d)}_{l}\rangle,\nonumber\\
V_1 &=& -m_i g^2\omega_i^2\left(R^2+\frac{m_a^2}{M^2}r^2+\frac{2m_a}{M}Rr\right)\label{eq_V1},\\
V_2 &=& -\frac{i g \omega_i m_i}{\hbar}\left(E_{l'}-E_{l}\right)\left(R^2+\frac{m_a^2}{M^2}r^2+\frac{2m_a}{M}Rr\right).\label{eq_V2}
\end{eqnarray}
\end{widetext}

\begin{figure*}
\includegraphics[width=16cm]{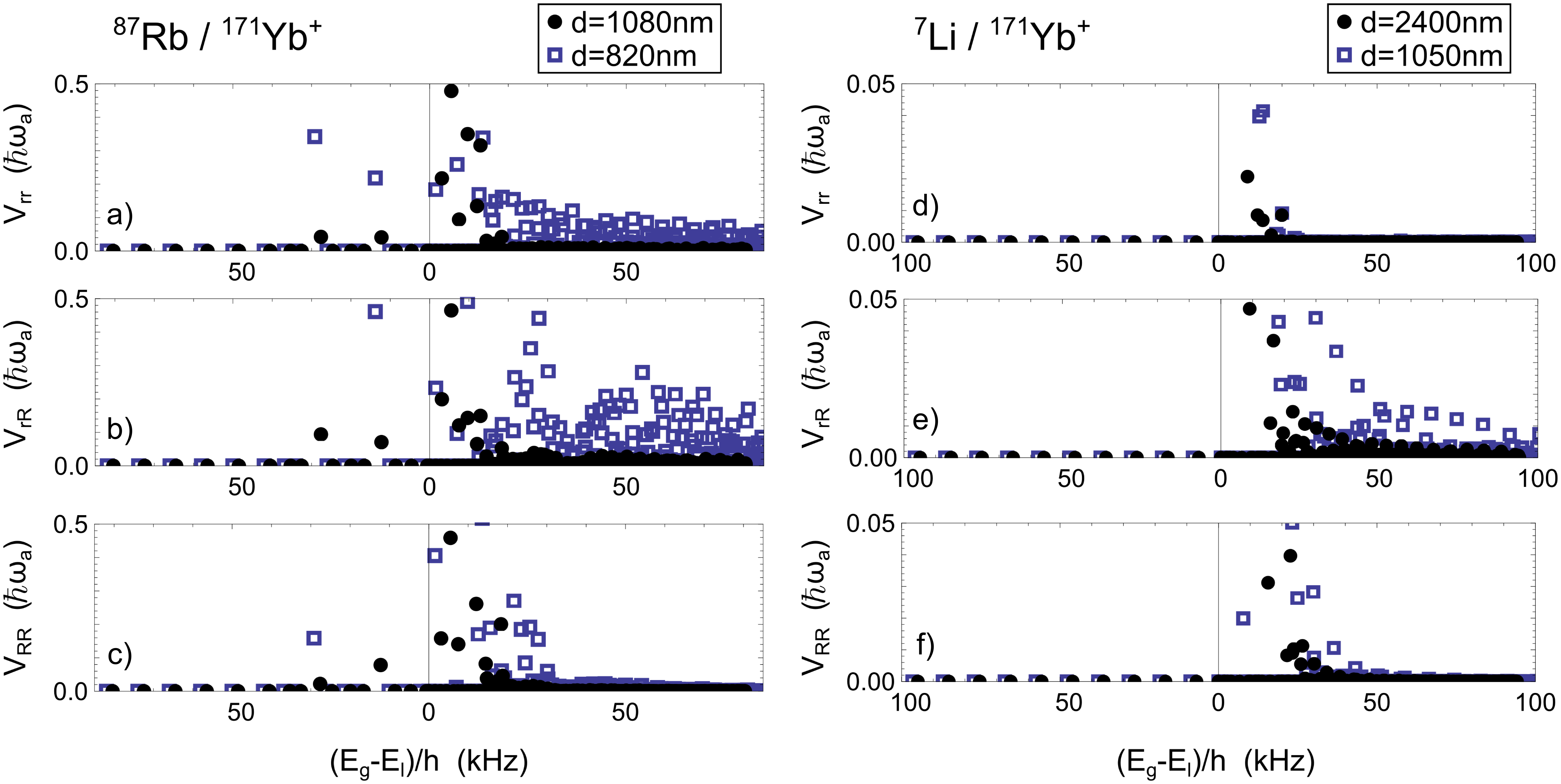}
\caption{\label{fig:matelems}
(Color online) a)-c): Matrix elements $V_{rr}$, $V_{rR}$ and $V_{RR}$ in units of $\hbar\omega_a$ for $^{87}$Rb and $^{171}$Yb$^+$ with $\phi_e=-\pi/4$, $\phi_o=\pi/4$, $\omega_a=2\pi\times 1.8$~kHz and $\omega_i=2\pi\times 9.9$~kHz for $d=1080$~nm and $d=820$~nm. The energy separation of the coupling state is plotted on the horizontal axis. Since stable ion trapping requires $\Omega\gtrsim 2\pi\times 30$~kHz (for $a=0$), only states with larger detunings than this can cause resonances. It can be seen that the elements $V_{rr}$ and $V_{rR}$ couple states that are separated in energy by more than 30~kHz, whereas $V_{RR}$ remains localized in energy such that it cannot cause resonant couplings. For inter-well distances of $\lesssim$~820~nm, the tunnelling rate reaches $J/h=58$~Hz but the matrix elements of $V_{rr}$ and $V_{rR}$ already reach 10\% of the atomic trapping frequency for energy differences around 85~kHz. Resonances with this strength would significantly affect the energy spectrum. This would render state preparation significantly more difficult and cause the two-mode approximation for the double-well to break down.  d)-f): Matrix elements for $^{7}$Li and $^{171}$Yb$^+$ for $d=2400$~nm and $d=1050$~nm and the same trap frequencies. At $d=1050$~nm, we find a tunnelling rate of $J/h=150$~Hz, whereas the micromotion induced couplings are at the percent level.
}
\end{figure*}

The larger these matrix elements are, the less we expect the secular approximation to hold. Additionally, the coupling between two states may become resonant for states belonging to different Floquet classes when $\epsilon_{j'l'}=\epsilon_{jl}$. We note that the selection rules for the coupling between different Floquet classes are given by $|j-j'|=1$ for $V_2$, and  $|j-j'|=2$ for $V_1$ [see also Eq.~(\ref{eq:genmatrxel})]. When the micromotion Hamiltonian cannot be considered a perturbation, the problem has to be solved by taking a large amount of Floquet classes into account. However, for large inter-well separation, we do not expect the micromotion to play a significant role as the atom is trapped too far away from the ion to sense the small amplitude micromotion of the ion. As $d$ is decreased, we expect the micromotion term to become increasingly important.

To quantify this, we will study how strongly the micromotion couples the states of interest $|\Phi_{e,g}^{(d)}\rangle$ to states with a different energy as $d$ is decreased. We limit the discussion to $|\Phi_{g}^{(d)}\rangle$ as it can be shown that the effects are very similar for $|\Phi_{e}^{(d)}\rangle$. The energy difference prefactor $(E_{l'}-E_{l})$ causes $V_2$ to be the main perturbing term~\cite{Nguyen:2012} and we focus on it from now on. Let us introduce the following notation for the (absolute value of) these matrix elements:

\begin{eqnarray}
V_{rr}&=&\frac{gm_i\omega_i\left(E_g-E_l\right)}{\hbar}\frac{m_a^2}{M^2}|\langle \Phi_g^{(d)}|r^2|\psi_l^{(d)}\rangle|,\nonumber\\
V_{rR}&=&\frac{gm_i\omega_i\left(E_g-E_l\right)}{\hbar}\frac{m_a}{M}|\langle \Phi^{(d)}_g|rR|\psi^{(d)}_l\rangle|,\nonumber\\
V_{RR}&=&\frac{gm_i\omega_i\left(E_g-E_l\right)}{\hbar}|\langle \Phi^{(d)}_g|R^2|\psi^{(d)}_l \rangle|.\nonumber
\end{eqnarray}

In Fig.~\ref{fig:matelems}~a)-c) the resulting matrix elements for the case discussed in Fig.~\ref{FIG:ionmovingspecwithandwithoutIA} are shown (i.e., for the atom-ion pair $^{87}$Rb/$^{171}$Yb$^+$). Here, the matrix elements are plotted for $d=1080$ and 820~nm and the detuning of the coupling state $E_g-E_l$. Clearly, the couplings are largest to states that are nearby in energy. For $V_{RR}$, the couplings quickly fall off to zero as the coupling state is further separated in energy. The terms involving $r$ and $r^2$, however, have significant couplings with states that are far separated in energy.  

This behaviour is a direct consequence of the non-linear interaction between the atom and the ion. In the center-of-mass coordinate the basis functions are Fock states and the matrix elements involving only $R$  have well-defined selection rules: $\langle f_{n'} |R| f_n\rangle \neq 0$ for $|n-n'| = 1$ and $\langle f_n |R^2| f_{n'}\rangle \neq 0$ for $|n-n'| = \{0,2\}$. Since the atomic and ionic ground state are only slightly perturbed by the atom-ion interaction, as it is clear from section~\ref{Sec_moving}, few Fock states are involved and due to the selection rules, only coupling to nearby states occur. Because there are also selection rules on $j$, the coupling states need to be of a different Floquet class. These states are unlikely to become resonant since $\Omega \gg \omega_R$. This situation is comparable to the effect that micromotion has on a single trapped ion. For the atom-ion scattering states in the relative coordinate, however, there are no selection rules owing to the non-linear atom-ion interaction. For instance, $\langle \Phi_k |r^2| \Phi_{k'}\rangle \neq 0$ in general for any $k$ and $k'$ that have the same symmetry. Thus, the states of interest couple to highly excited states belonging to different Floquet classes. 

For the case studied here, $\Omega=2\pi\times$~70.5~kHz, which corresponds to $q=0.4$. For inter-well distances of about 820~nm, the couplings to states that are $\sim$~70~kHz separated in energy - and can therefore become resonant - reach up to 10\% of the atomic trapping frequency. Such large couplings cause significant deviations from the secular solution if resonances occur. This situation is indeed quite similar to the case studied in Ref.~\cite{Nguyen:2012}. Hence, we expect a large number of energy levels crossing the ground states with increasing strength as $d$ is reduced. This will render the state preparation more difficult, when considering schemes where $d$ is slowly reduced to initiate atomic tunnelling. Only superb experimental control over $d(t)$ will allow the diabatic transfer over avoided crossings such that atomic excitations within the wells are prevented. 

It is interesting to note that the prefactors to the terms containing $r$ and $r^2$ in equations (\ref{eq_V1}) and (\ref{eq_V2}), that will cause the largest effects, are $m_a/M=\mu/m_i$ and $m_a^2/M^2$, respectively. This suggests that adverse micromotion effects may be reduced by choosing $m_i \gg m_a$, This observation is in line with the classical study of Cetina {\it et al.}~\cite{Cetina:2012} concerning the limits of atom-ion sympathetic cooling. As a comparison for the case of the double-well, we plot also the matrix elements for the case of $^7$Li and $^{171}$Yb$^+$ [see Fig.~\ref{fig:matelems}~d)-f)]. We see that the matrix elements are indeed smaller for this case and we reach a tunnelling rate of $J/h=150$~Hz with micromotion induced couplings on the percent level of the atomic trap frequency or $\sim$~20~Hz.

\begin{figure}
a)
\includegraphics[width=8cm]{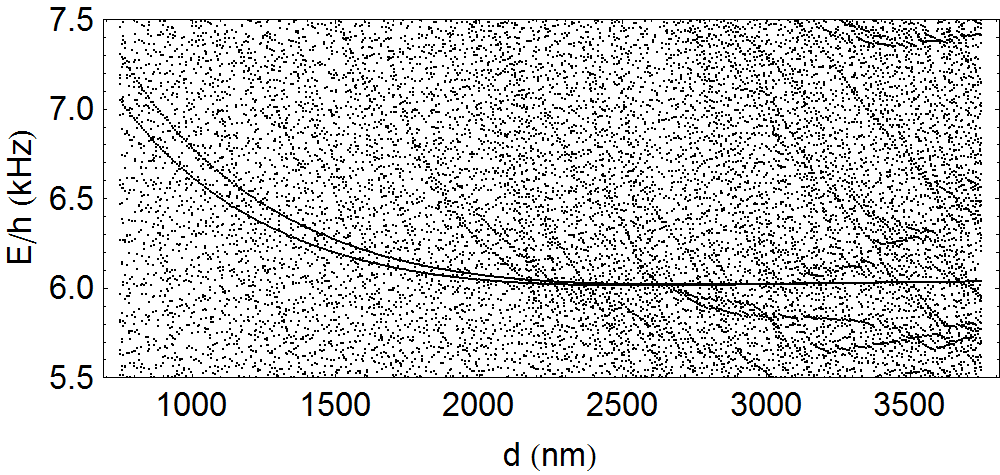}
b)
\includegraphics[width=8cm]{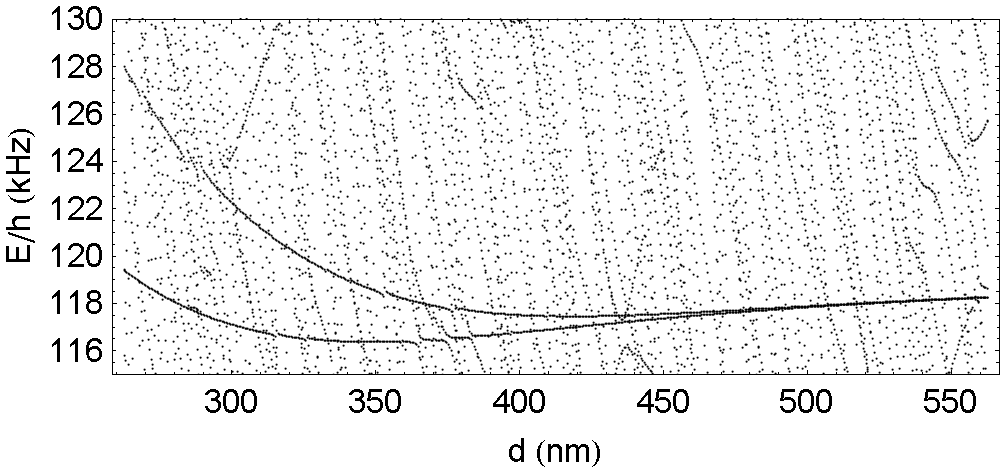}
\caption{\label{fig:mmLiYb}
 a) Spectrum around the states $|\Phi^{(d)}_{e,g}\rangle$ for $^7$Li and $^{171}$Yb$^+$ with $\omega_a=2\pi\times 1.8$~kHz, $\Omega=2\pi\times 70.5$~kHz for $q=0.4$ and $a=0$ such that $\omega_i\approx 2\pi\times 10$~kHz. For the calculation we took $j=-2,...,2$, $\phi_e=-\pi/4$, and $\phi_o=\pi/4$. To construct the basis 13500 states were used. The Hamiltonian was diagonalized for 500 different values of $d$. It can be seen that many energy crossings appear but the avoided crossings remain very small. b) Spectrum for $^7$Li and $^{171}$Yb$^+$ with $\omega_a=2\pi\times 98$~kHz, $\Omega=2\pi\times 967$~kHz for $q=0.4$ and $a=0$ such that $\omega_i\approx 2\pi$~137~kHz. For the calculation we took $j=-2,...,2$, $\phi_e=\pi/3$, and $\phi_o=-\pi/3$. To construct the basis 8640 states were used. The Hamiltonian was diagonalized for 600 different values of $d$. It can be seen that many energy crossings appear and that the avoided crossings become bigger as $d$ decreases.  
}
\end{figure}

Finally, we have numerically solved the full double-well problem including micromotion. The required total Hilbert space dimensions scale as $N_{com}\times N_{rel} \times N_{echos}$ - that is the number of center-of-mass states to be taken into account, times the number of relative states times the number of Floquet classes. Clearly, the calculation becomes very hard already at small $d$. For instance, calculating the eigenenergies for the case presented in Fig.~\ref{FIG:ionmovingspecwithandwithoutIA} taking only the classes $j=-2,..,2$ into account, would require a Hilbert space of dimension $32000\times 32000$. For the case of $^7$Li and $^{171}$Yb$^+$ of Fig.~\ref{FIG:ionmovingspecYbLi} the situation is better and we diagonalised the Hamiltonian for 500 values of $d$ taking $j=-2,..,2$, such that 13500 states are used to form the basis. The result is shown in Fig.~\ref{fig:mmLiYb} a). We can see that although many more energy levels are present in the spectrum, the coupling to the states $|\Phi^{(d)}_{e,g}\rangle$ is indeed very small, so that the micromotion should not pose a problem for the parameters considered here. 

As a second example, we calculate the spectrum around the ground states for a higher trap drive frequency. To reduce the numerical complexity we use $^7$Li and $^{171}$Yb$^+$ and set all energy scales to similar values, i.e. $\omega_a=2\pi\times 98$~kHz, $\Omega=2\pi\times 967$~kHz for $q=0.4$ and $a=0$ such that $\omega_i\approx 2\pi\times 137$~kHz, we obtain the eigenenergy spectrum shown in Fig.~\ref{fig:mmLiYb}. For this calculation we took $j=-2,...,2$, $\phi_e=\pi/3$, and $\phi_o=-\pi/3$. Thus, the Hilbert space comprised of 8640 basis states. As is illustrated in Fig.~\ref{fig:mmLiYb}, that also for this case, we see that the avoided crossings remain relatively small,  and therefore enabling us to apply the secular approximation for the double-well system at large enough separations $d$. Since the avoided crossings remain small, we do not expect that taking more Floquet classes into account will significantly alter the results for the values of $d$ plotted. On the other hand, for smaller $d$, the ground state presents many avoided crossings, and therefore more Floquet classes are required to reach convergence.

\section{Experimental implementation}
\label{sec:expimp}

\subsection{Optical potentials}

Atomic double-well potentials can be created by using optical tweezers or standing wave laser fields~\cite{Albiez:2005,Gati:2007,Levy:2007}. Such fields are sufficiently strong to also trap an ion, as it has first been shown in a recent proof-of-principle experiment~\cite{Schneider:2010}. An all optical or hybrid optical/Paul trap would solve the issues related to micromotion. In reference~\cite{Enderlein:2012}, for instance, a single ion was trapped in a standing wave laser field. It is feasible to alter this setup to allow both a harmonic potential for the ions and a multi-well potential for the atoms by overlapping trapping beams of suitable frequencies. Hence, the system studied in this paper may be realised by using optical trapping fields, or by means of a hybrid approach in which a Paul trap is combined with an optical one.

\subsection{Microtraps}

Atomic double-well systems have been created in atomic microtraps, or atom chips~\cite{ACbook11}, where atoms are magnetically trapped by means of integrated current carrying wires~\cite{Reichel:1999,Ott:2001,Folman:2002}. Planar ion traps have also become available in recent years~\cite{Seidelin:2006}, whose development was mainly driven by the prospect of developing a scalable quantum computer~\cite{Cirac:2000,Kielpinski:2002}. It seems feasible to combine the technologies of atom chips and planar ion traps to implement the proposed setup. Recently, for instance, magnetic field gradients were used in planar ion traps for spin detection or to design spin-spin interactions~\cite{Mintert:2001,Leibfried:2007,Ospelkaus:2011,Wang:2009,Welzel:2011}. Those traps combine electrodes on the surface of a chip that form the ionic trap with current carrying wires to produce magnetic gradient fields. Gradients ranging from 1-20~T/m at $\sim$100~$\mu$m above the surface can be reached. Slight modifications to such a setup would lead to magnetic field patterns that can trap atoms. For instance, we have calculated that by using a layered design, in which a planar ion trap is mounted on top of a current carrying structure, results in magnetic gradients of about 7~T/m. For atoms trapped 100~$\mu$m above the surface, this would correspond to trapping frequencies in the kHz range~\cite{Joger:2013} with appropriately aligned bias fields. Such a setup could be combined with radio-frequency fields as well, leading to adiabatic dressed double-well potentials, in analogy to the works of Refs.~\cite{Betz:2011,LeBlanc:2011}. In such a setup, the inter-well separation could be dynamically tuned by changing the frequency of the dressing field.

\section{Conclusions}
\label{sec:concl}

We have theoretically studied the dynamics of an atomic double-well system in the presence of a single trapped ion. We have found, under the assumption that the ion is not moving, that a one-dimensional calculation provides the same physical picture as a three-dimensional one. The spin of the ion can control the tunnelling rate via the state dependent short-range phase, as also discussed in reference~\cite{Gerritsma:2012}. When the ion is allowed to move, we find that the atomic tunnelling rate between the wells also couples to the ion motion. In this way, the tunnelling can be controlled by the motion of the ion. As the motional state of trapped ions is routinely engineered and read-out in state-of-the-art experiments~\cite{Leibfried:1996,Gerritsma:2010} by coupling it to its internal (spin) state, this may allow engineering atom-ion states in which the atomic position is entangled with the ion motion or spin. We have analyzed a scheme in which the inter-well distance is dynamically reduced to allow the atom to tunnel depending on the motional state of the ion. Imperfect ion ground state cooling will result in reduced tunnelling contrast. Since ion heating may occur on the timescale of the tunnelling dynamics in experiments where the ion is trapped close to the electrodes, we plan to analyze the effect of ion heating and cooling on the tunnelling dynamics in the future. The coupling between atomic tunnelling and ion motion can also be seen as a unit-cell for a larger atom-ion quantum simulator, in which the atomic dynamics is coupled to phonons in an ion crystal~\cite{Bissbort:2013}. 

We have also analyzed the effect of micromotion on the energy spectrum of the double-well system. The micromotion causes many extra avoided crossings in the spectrum as coupling to states belonging to different Floquet classes becomes possible for small inter-well separation $d$. The exact strength of these crossings depends on the trap parameters, but we conclude that it is a good idea to choose an ion-atom combination with a large mass ratio. The avoided crossings will cause trouble in state preparation as the wells have to be brought together diabatically with respect to the crossings, but without exciting the atoms to higher trap states. Additionally, the two-mode approximation may break down in a many-body scenario when including the micromotion, complicating its theoretical description. 

In linear Paul traps, the dynamical electric field is only used to confine the ions in two of the three directions and it may be advisable to have the double-well separation in the third direction where the ion is confined by static fields. Nonetheless, due to the spherical symmetry of the atom-ion interaction potential, there will be some minimum distance where the secular approximation will fail~\cite{Nguyen:2012}. 

We have given two possible routes towards experimental implementation of the double-well system. Either the double-well potential is created by using magnetic fields, or with optical fields. Both technologies are compatible with ion trapping, putting an experimental realization of the considered system within reach. 

\acknowledgements

This work was supported by the ERC (grant 337638) (R.G. and J.J.), by an internal grant of the University Mainz (Project `Hybrid atom-ion micortrap') (R.G. and J.J.), and by the excellence cluster `The Hamburg Centre for Ultrafast Imaging - Structure, Dynamics and Control of Matter at the Atomic Scale' of the Deutsche Forschungsgemeinschaft (A.N.). The authors acknowledge useful discussions with Andrew Grier and Krzysztof Jachymski, and are grateful to Ferdinand Schmidt-Kaler, Tommaso Calarco, and Zbigniew Idziaszek for previous work on a related project. RG thanks Ulrich Poschinger for proof reading the manuscript.

\appendix
\section{3D calculation}\label{app_3D_M}

The procedure for obtaining the 3D eigenenergies and states has been described in Ref.~\cite{Gerritsma:2012} and for completeness we give more details here. The 3D Hamiltonian is given by $H_{3D}=H_{3D}^{(0)}+H_{3D}^{(1)}$, where

\begin{eqnarray}\label{eq_3D}
H_{3D}^{(0)}&=&-\frac{\hbar^2\nabla_a^2}{2m_a}+\frac{1}{2}m_a\omega_{\perp}^2 r_a^2-\frac{C_4}{r_a^4},\\
H_{3D}^{(1)}&=&V_{dw}(z_a)-\frac{1}{2}m_a\omega_{\perp}^2 z_a^2.
\end{eqnarray}

\noindent Here $\omega_{\perp}$ is the trapping frequency in the transverse direction. We introduce spherical coordinates $(r_a,\theta_a,\phi_a)$, where we kept the subscript $a$ for denoting the coordinates that belong to the atom. The solutions to the radially symmetric equation~(\ref{eq_3D}) are of the form $\bm{\psi}^0_{nlm}(\mathbf{r}_a)=Y_l^m(\theta_a,\phi_a) \psi^0_{nl}(r_a)/r_a$, where $Y_l^m(\theta_a,\phi_a)$ are spherical harmonics, and $\psi^0_{nl}(r_a)$ are the solutions to the radial Schr\"odinger equation. Since the potential does not depend on the azimuthal direction, the quantum number $m$ is conserved and for simplicity we set $m=0$. 

The radial Schr\"odinger equation is given by:

\beq\label{eqspherical}
E_{nl}^{(0)}\psi^0_{nl}\!=\!\left(\!-\frac{\hbar^2}{2m_a}\frac{\partial^2}{\partial r_a^2}+\frac{\hbar^2 l(l+1)}{2m_ar_a^2}+\frac{m_a\omega_{\perp}^2 r_a^2}{2}-\frac{C_4}{r_a^4}\right)\!\psi^0_{nl}.\nonumber
\eeq
To numerically solve this equation, we make use of the renormalized Numerov method~\cite{Johnson:1977} with 

\beq
\tilde{\psi}^0_{nl}(r_a)\propto \sqrt{r} \left[J_{l+1/2}(\xi)+\tan(\delta)Y_{l+1/2}(\xi)\right]
\eeq
\noindent as a boundary condition, where $\xi=\sqrt{m_a/\mu} \, R^*/r_a$. The mixing angle $\delta$ is related to the 3D short-range phase as: $\delta=-\phi-l\pi/2$. We then expanded the solution of the full Hamiltonian $H_{3D}$ onto the solutions of Eq.~(\ref{eq_3D}), namely $\bm{\psi}^{(d)}(\mathbf{r_a})=\sum_{k} c_{k} \bm{\psi}^0_{k}(\mathbf{r_a})$, where $k$ denotes the pair of quantum numbers $(n,l)$.  Thus, both the wavefunctions and the corresponding energies are obtained by diagonalising the Hamiltonian $H_{3D}$, whose matrix elements are given by:
\begin{eqnarray}
H_{kk'}\!&\!=\!&\!\left(\!E^{(0)}_{k}\!+\!\frac{m_a\omega_a^2d^2}{8}\,\right)\delta_{kk'}\!+\frac{m_a\omega_a^2}{4}\left(\!\frac{\mathcal{M}^{(4)}_{kk'}}{2d^2}-3 \mathcal{M}^{(2)}_{kk'}\!\right),\nonumber\\
\mathcal{M}^{(j)}_{kk'}\!&\!=\!&\!\int d\mathbf{r}_a^3 \bm{\psi}^{0*}_{k'}(\mathbf{r}_a) \,\cos^j(\theta_a)\, r_a^{j} \,\bm{\psi}^{0}_k(\mathbf{r}_a),\nonumber
\end{eqnarray}

\noindent with $j=2,4$. These matrix elements can be explicitly written as

\begin{eqnarray}
\mathcal{M}^{(j)}_{nl,n'l'}\!&\!=\!&\!\int_0^{\infty} \psi_{n'l'}^{0*}(r_a) r_a^{j}\psi_{nl}^{0}(r_a) dr_a\nonumber\\
\!&&\!\times2\pi\int_0^{\pi}\sin\theta_a\,\cos^j\!\theta_a \, Y^*_{l'}(\theta_a)Y_l(\theta_a) d\theta_a,\nonumber
\end{eqnarray}
whose determination relies on the computation of the following Clebsch-Gordon coefficients:

\begin{eqnarray}
C^{(j)}_{ll'}&\!=\!&2\pi\int_0^{\pi}\sin\theta_a\,\cos^j\!\theta_a \, Y^*_{l'}(\theta_a)Y_l(\theta_a) d\theta_a,\nonumber\\
C^{(2)}_{l,l}&\!=\!&\frac{2l(l+1)-1}{4l(l+1)-3},\nonumber\\
C^{(2)}_{l,l+2}&\!=\!&\frac{(l+1)(l+2)\sqrt{5+4l(l+3)}}{(2l+1)(2l+3)(2l+5)},\nonumber\\
C^{(4)}_{l,l}&\!=\!&\frac{3[3+2l(l+1)(l^2+l-4)]}{(2l-3)(2l-1)(2l+3)(2l+5)},\nonumber\\
C^{(4)}_{l,l+2}&\!=\!&\frac{2(l+1)(l+2)[-3+2l(l+3)]\sqrt{5+4l(l+3)}}{(2l-1)(2l+1)(2l+3)(2l+5)(2l+7)},\nonumber\\
C^{(4)}_{l,l+4}&\!=\!&\frac{(l+1)(l+2)(l+3)(l+4)\sqrt{9+4l(l+5)}}{(2l+1)(2l+3)(2l+5)(2l+7)(2l+9)}.\nonumber
\end{eqnarray}

\noindent All other coefficients vanish due to selection rules on the quantum number $l$. Besides this, we note that the coefficient matrices are symmetric $C^{(j)}_{ll'}=C^{(j)}_{l'l}$.

\begin{figure}
\includegraphics[width=8cm]{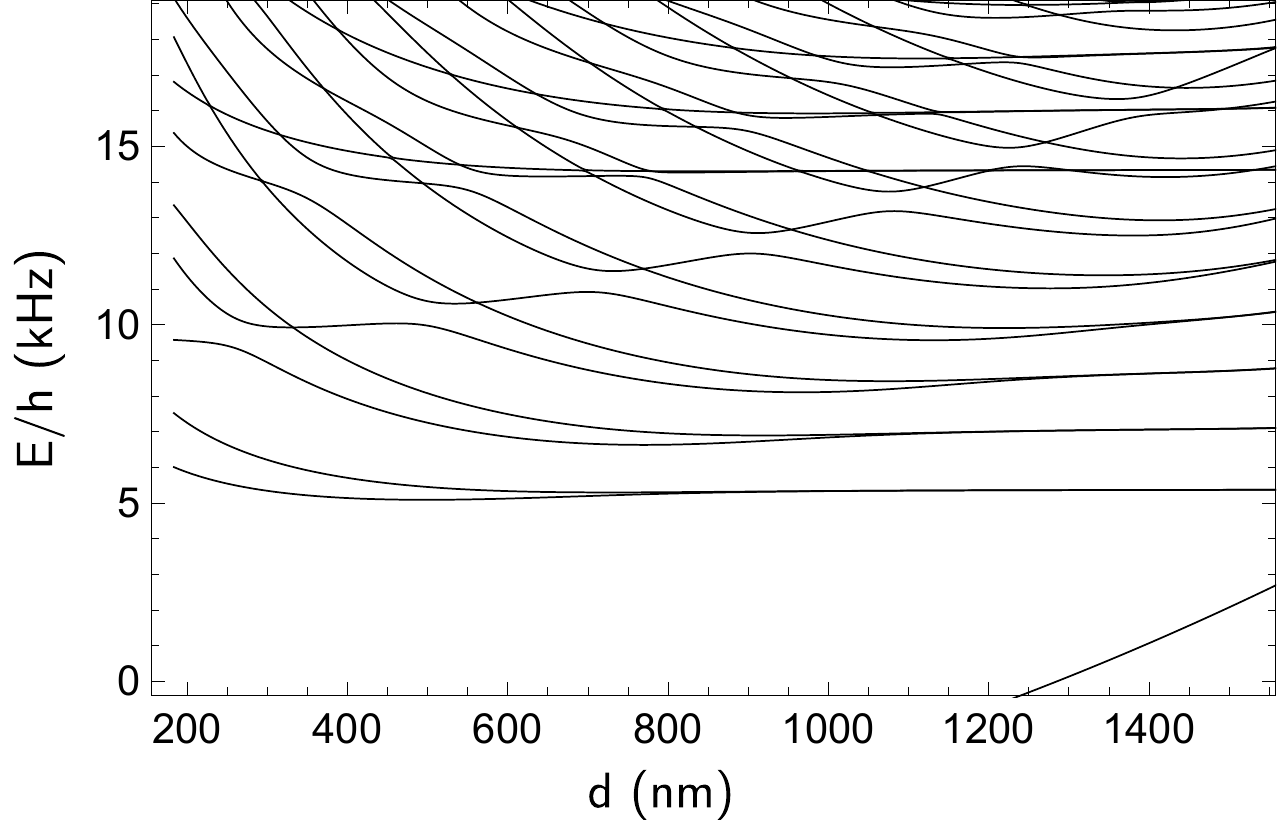}\caption{\label{FIG:atomonlyspec3D} Three-dimensional eigenenergy spectrum as a function of inter-well separation $d$ for a $^{87}$Rb atom and a pinned $^{171}$Yb$^+$ ion assuming a 3D short range phase of $\phi=\pi/4$ and $\omega_a=2\pi\times 1.8$~kHz, $\omega_{\perp}=2\pi\times 4.5$~kHz .  
}
\end{figure}

 In Fig.~\ref{FIG:atomonlyspec3D} we show the spectrum for the parameters $\phi=\pi/4$,  $\omega_{\perp}=2\pi\times 4.5$~kHz, and $\omega_a=2\pi\times 1.8$~kHz, which is the same axial trapping frequency as used in Fig.~\ref{FIG:atomonlyspec}. We note that for such calculation we have obtained a set of 1838 eigenfunctions and energies including bound states as well as trap states~\cite{Doerk:2010,Gerritsma:2012}. Apart from the extra angular integration, the whole procedure is very similar to the one used in the 1D calculation and the resulting energy spectrum looks indeed very similar. There are, however, some differences. First of all, since there are more degrees of freedom, more levels appear. Note also that not all levels are plotted since we limited our discussion to $m=0$. Secondly, the exact way in which the energy levels split close to the ion is different. This behavior, however, is  related to the choice of the short-range phase. To determine the exact form experimental input would be needed in the form of a scattering length.

\section{Micromotion Hamiltonian}\label{App_mm}

Following Refs.~\cite{Cook:1985,Nguyen:2012} we start by writing the ion wavefunction as:

\beq
\Psi(z_i,t)={\rm exp}\left(-\frac{i}{4\hbar}m_i q \Omega z_i^2\sin (\Omega t)\right)w (z_i,t).
\eeq

\noindent By replacing this function into Eq.~(\ref{H_ion}) the following effective Hamiltonian for the wavefunction $w (z_i,t)$ is obtained

\beq
H_{eff}(t)=\frac{p_i^2}{2m_i}+\frac{1}{2}m_i\omega_i^2z_i^2+H_{mm}(t)
\eeq

\noindent with the micromotion term given by:

\beq
H_{mm}(t)=-m_ig^2\omega_i^2z_i^2\cos(2\Omega t)-g\omega_i \{z_i,p_i\}\sin(\Omega t).
\eeq

\noindent The secular trapping frequency is given by:

\beq
\omega_i=\frac{\Omega}{2}\sqrt{a+\frac{q^2}{2}}.
\eeq

In order to evaluate the matrix elements of $H_{mm}(t)$, that is, Eqs.~(\ref{eq_V1},\ref{eq_V2}), it is very useful to note that $\langle \psi^{(d)}_{l'} |\{p,r\}|\psi^{(d)}_l\rangle = i \mu\langle \psi^{(d)}_{l'} |[H^{(d)},r^2]|\psi^{(d)}_l\rangle/\hbar$. Similarly, $\langle \psi^{(d)}_{l'} |p|\psi^{(d)}_l\rangle = i \mu\langle \psi^{(d)}_{l'} |[H^{(d)},r]|\psi^{(d)}_l\rangle/\hbar$~\cite{Nguyen:2012}. Equivalent equations hold for the center-of-mass coordinate too.


\end{document}